\documentstyle[preprint,prb,aps]{revtex}

\begin{document}

\draft
\title{Transverse Magnetoresistance of GaAs/AlGaAs Heterojunctions in
the Presence of Parallel Magnetic Fields}

\author{J.M. Heisz and E. Zaremba}
\address{Department of Physics,
Queen's University,
Kingston, Ont., Canada K7L 3N6}

\date{\today}

\maketitle

\begin{abstract}
We have calculated the resistivity of a GaAs\slash AlGaAs
heterojunction in the presence of both an in--plane magnetic
field and a weak perpendicular component using a 
semiclassical Boltzmann transport theory. These calculations
take into account fully the distortion of the Fermi contour
which is induced by the parallel magnetic field.
The scattering of electrons is assumed to be due
to remote ionized impurities. A positive magnetoresistance is
found as a function of the perpendicular component, in good
qualitative agreement with experimental observations. The main
source of this effect is the strong variation of the electronic
scattering rate around the Fermi contour which is associated
with the variation in the mean distance of the electronic states
from the remote impurities. The magnitude of the 
positive magnetoresistance is strongly correlated with
the residual acceptor impurity density in the GaAs layer. 
The carrier lifetime anisotropy also
leads to an observable anisotropy in the resistivity with
respect to the angle between the current and the direction of
the in--plane magnetic field.
\end{abstract}


\section{Introduction}

Magnetotransport in a two-dimensional electron gas (2DEG) is an
extremely rich and complex subject. In the more usual geometry
with a magnetic field perpendicular to the plane of the 2DEG, the
magnetoresistance exhibits Shubnikov-de Haas oscillations which
are the low-field precursors of the quantum Hall effect. This
behaviour is a consequence of the quantization of the in-plane
orbital motion into Landau levels.
The situation with a magnetic field in the plane of the 2DEG
is quite different since the cyclotron motion now competes with
the effects of the potential confining the electrons. This leads
to an interesting modification of the electronic energy band
structure which was recently analyzed in detail for typical
GaAs/AlGaAs heterojunctions\cite{mine,SmJu93}. These
calculations account for the observed
depopulation of higher subbands\cite{fletch} which occurs with
increasing parallel magnetic field.

In comparison to the perpendicular field geometry,
transport in the presence of a parallel field has
received relatively little attention. One of the earliest
studies is that of Englert et al.\cite{englert} who observed a
pronounced positive transverse magnetoresistance when an
in--plane magnetic field was tilted slightly out of the plane
of the 2DEG. This interesting observation was not understood at
the time. More recently, the same effect was studied in
considerably more detail by Leadley et al.\cite{leadley}
for a number of GaAs/AlGaAs heterojunctions. The explanation of
this effect is the primary focus of our work.

A positive transverse magnetoresistance is known to occur in
situations where conduction is provided by different kinds of
carriers, for example, a heterojunction with two or more
occupied subbands. In accord with classical theory\cite{ziman},
samples with only a single occupied subband do not show the
effect (in fact, a negative magnetoresistance attributable to
weak localization is usually observed). However the situation
changes when the system is subjected to a parallel magnetic
field which  affects the electronic structure and leads to a 
field-induced anisotropy of the 2D Fermi contour.
As noted by Leadley et al.\cite{leadley}, this is analogous
to the Fermi surface anisotropy found in metals and
should similarly lead to a positive magnetoresistance\cite{pipard}.
Although the anisotropy of the
Fermi contour is important, we shall see that it is not the only
effect contributing to the positive magnetoresistance.

There have been a few theoretical considerations of this problem.
The first calculation of transport in the presence of a parallel
magnetic field was carried out by Tang and
Butcher\cite{tang1,tang2} who considered a model in
which the 2DEG is confined by a harmonic potential.
The electronic states and energy dispersion can be obtained 
analytically for this model which simplifies the solution
of the transport problem. They further assume that
the electrons are scattered by short range $\delta$-function
potentials and demonstrate within the framework of the Boltzmann
transport theory that the conductivity of the electron gas is
anisotropic with respect to the angle between the current and
the parallel magnetic field. However they do not consider the
additional effects of a perpendicular magnetic field. 
A similar model was used by Smr\v{c}ka\cite{smrcka} in an
attempt to explain the observed positive magnetoresistance.
However his calculation invokes a dc
conductivity in the third dimension perpendicular to
the plane of the 2DEG which has no obvious physical meaning.  
More recently, St\v{r}eda et al.\cite{Streda} developed a
semiclassical Boltzmann transport theory of the
magnetoresistance within a constant relaxation time
approximation. The theory assumes that currents flowing in a
direction perpendicular to the in--plane magnetic field induce a
transverse Hall field perpendicular to the 2DEG as first
suggested by Stern\cite{stern}. It is then argued that this field in
turn leads to a correction to the in--plane conductivity  which
is the source of the positive magnetoresistance. However, it is
unclear how this mechanism might account for the
magnetoresistance when the current is flowing {\it parallel} to
the in--plane field. In any event, since the argumentation is
rather heuristic, a more rigorous explanation of the experimental
observations is needed. We have therefore
performed a more thorough calculation of the
transport properties in a tilted magnetic field, utilizing a
realistic model of both the heterojunction electronic structure
and the scattering of electrons from remote ionized impurities.
We shall demonstrate that the inherent anisotropy of these
properties is ultimately the source
of the positive transverse magnetoresistance.

\vfil\break
We perform our calculations for model heterojunctions which
correspond to the G590 series of samples studied by 
Leadley et al.\cite{leadley} 
By modelling these particular samples we hope
to minimize any differences in electronic structure which might
otherwise obscure a comparison between theory and experiment.
The design characteristics of
these samples have been discussed elsewhere\cite{fletch}.  The
important variables are the areal acceptor and donor
densities ($n_A$ and $n_D$), which can be varied experimentally 
by means of the persistent photoconductivity effect.  
We shall deal specifically with two combinations, herein referred 
to as Sample~1 and Sample~2.  Sample~1 has an acceptor density of
$n_A=1.03\times 10^{11} cm^{-2}$ and a donor density of 
$n_D=4.53\times 10^{11} cm^{-2}$, giving a total electron density of
$n_T=3.5\times 10^{11} cm^{-2}$.  It corresponds to the
unilluminated experimental sample referred to as G590a which
has only one occupied subband. For Sample~2, the areal densities are
$n_A=0.1\times 10^{11} cm^{-2}$ and $n_D=7.2\times 10^{11} cm^{-2}$,
resulting in a total electron density of 
$n_T=7.1\times 10^{11} cm^{-2}$,
corresponding to the experimental sample referred to as G590c.
This sample is obtained from G590a by illumination and has
two occupied subbands.  A comparison of the theoretical 
and experimental zero-field mobilities provides one measure 
of how well we have been able to model these samples.
We find $20.6 m^2/Vs$ and $37.9 m^2/Vs$ for Sample~1 and
2, respectively, which compare favourably with the experimental
values of $17.4 m^2/Vs$ and $94.0 m^2/Vs$. We are therefore
confident that our model heterojunctions closely represent the 
actual experimental samples.
\bigskip

\section{Theory}

\subsection{Boltzmann Transport Theory}

In this section, we develop the Boltzmann transport equation for a
2DEG in the presence of a strong parallel magnetic field 
in the $x$-direction and a weak perpendicular component.
The parallel field is taken into account quantum
mechanically in the determination of the subband electronic 
structure (see Ref.\ \onlinecite{mine}).
As a function of the in--plane wavevector ${\bf k}$, 
the dispersion of the subband
energies $E_{n \bf k}$ is anisotropic and leads to noncircular constant
energy contours.  An example of this behaviour is shown in 
Figure~1, where the contours are illustrated for Sample 2
for two parallel magnetic fields, one ($B = 5T$) for which two
subbands are occupied and a second ($B = 10T$) for which only a
single subband is occupied.

The dynamics of the electrons in the presence of a perpendicular
magnetic field component $B_{\perp}$ will be described semiclassically.
The transport properties arising from an external electric field
{\bf E} are then determined using the Boltzmann equation
\begin{equation}
\left. {{\partial f_n({\bf k})}\over{\partial t}} \right\vert_{scat} = 
- \left ( {{e}\over{\hbar}} \right ) {\bf E} \cdot \nabla_{\bf k}
f_n ({\bf k}) - {{e}\over{\hbar c}} \left ( {\bf v}_{n{\bf k}} 
\times {\bf B}_{\perp} \right ) \cdot \nabla_{\bf k} f_n ({\bf k})
\end{equation}
where $f_n({\bf k})$ is the 
nonequilibrium distribution function of subband $n$ and
${\bf v}_{n{\bf k}} = \nabla_{\bf k} E_{n \bf k} / \hbar$ is the group 
velocity. Since the energy bands are neither parabolic nor
isotropic in the presence of a parallel magnetic field, the
electron velocity is not parallel to the wave vector ${\bf k}$.
Expressing $f_n({\bf k})$ as
\begin{equation}
f_n({\bf k}) = f^0(E_{n \bf k}) + g_n({\bf k})
\label{disfn}
\end{equation}
where $f^0(E_{n \bf k})$ is the equilibrium Fermi distribution and
$g_n({\bf k})$ is the deviation from equilibrium, the linearized
Boltzmann equation is given by
\begin{equation}
\left. {{\partial g_n({\bf k})}\over{\partial t}} \right\vert_{scat} = 
- \left ( {{e}\over{\hbar}} \right ) {\bf E} \cdot \nabla_{\bf k}
f^0 (E_{n \bf k}) - {{e}\over{\hbar c}} \left ( {\bf v}_{n{\bf k}} 
\times {\bf B}_{\perp} \right ) \cdot \nabla_{\bf k} g_n ({\bf k})\,.
\label{linear}
\end{equation}
The scattering term in (\ref{linear}) is
\begin{equation}
\left. {{\partial g_n({\bf k})}\over{\partial t}} \right\vert_{scat} = 
\sum_{n' {\bf k}'} \left [ g_{n'}({\bf k}') - g_{n}({\bf k}) \right ]
\; w_{n n'} ({\bf k}, {\bf k}')
\end{equation}
where $w_{n n'} ({\bf k}, {\bf k}')$ is the transition rate between the
states $n{\bf k}$ and $n'{\bf k}'$. It satisfies the detailed
balance condition $w_{n' n} ({\bf k'}, {\bf k}) = w_{n n'} ({\bf
k}, {\bf k}')$ which ensures the overall conservation of particle
number.

In the following we shall assume
that the scattering is due to impurities. In this
case, Fermi's golden rule gives the transition probability  
\begin{equation}
w_{n n'}({\bf k}, {\bf k}') = {{2 \pi}\over{\hbar}} \;
\overline{\vert \langle n{\bf k} \vert V({\bf r}) \vert n'{\bf k}'
\rangle \vert^2} \; \delta (E_{n{\bf k}} - E_{n'{\bf k}'} )
\end{equation}
where $V({\bf r})$ is the screened impurity potential. The bar
over the matrix element denotes an average over all possible
impurity configurations. The important property for the present
purposes is the energy-conserving delta function which corresponds 
to the elastic nature of the scattering. Making use of this
dependence and defining the quantity $\xi_n({\bf k})$ by
\begin{equation}
g_n({\bf k}) = {{\partial f^0(E_{n \bf k})}\over{\partial E_{n \bf k}}}
\xi_n({\bf k})
\end{equation}
equation (\ref{linear}) can be reduced to 
\begin{equation}
\tau_n^{-1} ({\bf k}) \; \xi_n({\bf k}) = \sum_{n' {\bf k}'}
w_{n n'} ({\bf k}, {\bf k}') \; \xi_{n'}({\bf k}')
+ e {\bf E} \cdot {\bf v}_{n {\bf k}}
+ {{e}\over{\hbar c}} \left ( {\bf v}_{n{\bf k}} 
\times {\bf B}_{\perp} \right ) \cdot \nabla_{\bf k} \xi_n ({\bf k})
\label{reduced}
\end{equation}
Here, $\tau_n({\bf k})$ is the quantum lifetime
\begin{equation}
\tau_n^{-1}({\bf k}) = \sum_{n' {\bf k}'}
w_{n n'} ({\bf k}, {\bf k}')
\label{lifetime}
\end{equation}
which represents the total unweighted probability of scattering from the
initial state $n{\bf k}$ to all available final states,
including states in other subbands. Due to the more
complicated energy subband structure in the presence of a parallel
magnetic field, $w_{n n'} ({\bf k}, {\bf k}')$ is not simply a
function of ${\bf k}- {\bf k}'$ and as a result, 
$\tau_n({\bf k})$ is in general an anisotropic function
of the wavevector {\bf k}.

The anisotropic energy band structure also complicates the
solution of the Boltzmann equation.
We find it useful to transform from the 
{\bf k}--space variables to a curvilinear coordinate system defined by
the energy variable $\epsilon \equiv E_{n{\bf k}}$ 
and an angular variable $\phi$
corresponding to the orientation of the vector {\bf k}.  To be specific,
the position in {\bf k}--space of a state in the $n$th subband is
referred to the position of the {\it minimum} of the subband energy  
which is displaced from $k_y = 0$ because of the parallel magnetic 
field. This new origin is implicitly assumed in the following. 
The definition of these variables is illustrated in Figure~1.
We have found that the use of the variable $\phi$ is more convenient 
in the present context than the phase angle $\theta$ 
which naturally appears in discussions of cyclotron 
motion (Ref.\ \onlinecite{ziman}, Chap. 9).

With this variable transformation, equation (\ref{reduced}) becomes
\begin{eqnarray}
\tau_n^{-1} (\epsilon, \phi) \; \xi_n(\epsilon, \phi) = 
\sum_{n'} {1\over2\pi} \int_0^{2\pi} && d\phi
J_{n'}(\epsilon, \phi') \; P_{n n'} (\epsilon, \phi, \phi') \;
\xi_{n'} (\epsilon, \phi') \nonumber \\
&&+ e {\bf E} \cdot {\bf v}_n ({\epsilon, \phi})
- \left ( {{eB_{\perp}}\over{\hbar c}} \right ) v_n ({\epsilon, \phi})
\hat{\bf t}_n \cdot \nabla_{\bf k} \xi_n(\epsilon, \phi)
\label{transformed}
\end{eqnarray}
where $\hat{\bf t}_n$ is a unit vector pointing in the direction
of cyclotron motion on the constant energy contour of the
$n$th subband at the point ${\bf k}$. The angular transition rate at
energy $\epsilon$ appearing in (\ref{transformed}) is defined by 
\begin{equation}
P_{n n'} (E_{n {\bf k}}, \phi, \phi') =
A{{m^*}\over{\hbar^3}} \;
\overline{\vert \langle n{\bf k} \vert V({\bf r}) \vert n'{\bf k}'
\rangle \vert^2} 
\label{scatrate}
\end{equation}
in terms of which the quantum lifetime is given by
\begin{equation}
\tau_n^{-1}(\epsilon, \phi) = \sum_{n'} {1\over 2\pi}
\int_0^{2\pi} d\phi' J_{n'} (\epsilon, \phi')
P_{n n'} (\epsilon, \phi, \phi')\,.
\label{lifetime2}
\end{equation}
The integrals in (\ref{transformed}) and (\ref{lifetime2})
contain the quantity
\begin{equation}
J_n(\epsilon, \phi)=
{{\hbar^2 k}\over{m^* \vert \nabla_{\bf k} E_{n {\bf k}} 
\cdot \hat{\bf k} \vert}}
\label{jacobian}
\end{equation}
which is a dimensionless form of the Jacobian of the variable
transformation. In this form it is just the ratio of the
free particle velocity $\hbar k / m^*$ to the component of the actual
velocity ${\bf v}_n({\bf k})$ in the direction $\hat{\bf k}$. The
deviation of this function from unity reflects the asymmetry of the
constant energy contours in the presence of a parallel magnetic field.

The solution of (\ref{transformed}) is easily obtained 
using a Fourier expansion in the angular variable, i.e.
\begin{equation}
\xi_n(\epsilon, \phi) = \sum_m a_n^{(m)} (\epsilon) \; e^{im\phi}\,.
\end{equation}
Multiplying (\ref{transformed}) by $e^{-im\phi}J_n(\epsilon, \phi)$ 
and integrating over $\phi$ yields the set of linear equations
\begin{equation}
\sum_{m' n'} \left \lbrack A_n^{(m-m')} \delta_{nn'} -
B_{n n'}^{(m,-m')} +i m \omega_c^0
\delta_{mm'} \delta_{nn'} \right\rbrack a_{n'}^{(m')}
=b_n^{(m)}
\label{linears}
\end{equation}
with
\begin{equation}
A_n^{(m)}={1\over 2\pi}\int_0^{2\pi} d\phi 
e^{-im\phi} J_n(\phi) \tau_n^{-1}(\phi),
\end{equation}
\begin{equation}
B_{nn'}^{(m,m')}={1\over 2\pi} \int_0^{2\pi} d\phi e^{-im\phi} 
J_n(\phi) {1\over 2\pi} \int_0^{2\pi} d\phi' e^{-im'\phi'} 
J_{n'}(\phi') P_{nn'} (\phi, \phi')  
\label{matrix}
\end{equation}
and
\begin{equation}
b_n^{(m)}={1\over 2\pi} \int_0^{2\pi} d\phi e^{-im\phi} 
J_n(\phi) e{\bf E} \cdot 
{\bf v}_n(\phi)
\label{inhomog}
\end{equation}
In these equations, the dependence on the energy variable $\epsilon$ has
been suppressed.  We note that the free electron cyclotron frequency
$\omega_c^0={{eB_{\perp}}\over{m^* c}}$ appears in equation 
(\ref{linears}), and not the actual cyclotron frequency defined by 
$\omega_{cn}^{-1}= {{\hbar c}\over{eB_{\perp}}} \oint {{dk}\over{v_n}}$,
where the line integral extends over a constant energy contour.  

In the
absence of a parallel magnetic field the energy dispersion is isotropic,
$J_n(\phi)$ reduces to unity and the angular transition rate is only a
function of the difference $\phi-\phi'$.  As a result,
$\tau_n^{-1}(\phi)$ becomes independent of $\phi$, $b_n^{(m)}$ is
proportional to $\delta_{m, \pm1}$ and $a_n^{(\pm 1)}$ are
the only non--vanishing expansion coefficients. In this limit, one
recovers the usual form of the multisubband transport 
equations\cite{zaremba}.  

In the present situation however, 
the equations defining the Fourier expansion coefficients
are coupled. In practice, the expansion is truncated at some
finite number of terms and the set of equations in (\ref{linears})
is solved with the inhomogeneous term taking on two possible
values, one corresponding to an electric field in the 
$x$-direction and one in the $y$-direction.
Once these solutions are known, the 
transport current density is obtained from
\begin{eqnarray}
{\bf J} & = & -{2e\over A} \sum_{n{\bf k}} g_n({\bf k}) 
{\bf v}_{n{\bf k}} \nonumber \\
&=&{{m^* e}\over {2\pi^2\hbar^2}} \; \sum_{mn}
\int_{-\infty}^{\infty} d\epsilon
\left ( - {{\partial f^0(\epsilon)}\over{\partial \epsilon}}
\right ) 
a_n^{(m)}(\epsilon)
\int_0^{2\pi} d\phi J_n(\epsilon, \phi) {\bf v}_n(\epsilon, \phi) 
e^{im\phi}\,.
\end{eqnarray}
Since the Fourier coefficient $a_n^{(m)}$ depends linearly on the
electric field, we can define the conductivity tensor as
\begin{equation}
\sigma_{\mu\nu} = {{m^* e}\over {2\pi^2\hbar^2}} \; \sum_{mn}
\int_{-\infty}^{\infty} d\epsilon
\left ( - {{\partial f^0(\epsilon)}\over{\partial \epsilon}}
\right ) 
{\partial a_n^{(m)}(\epsilon) \over \partial E_\nu}
\int_0^{2\pi} d\phi J_n(\epsilon, \phi)  v_{n\mu}(\epsilon, \phi) 
e^{im\phi}\,.
\end{equation}
In the $B_\perp \to 0$ limit, one can show that $\sigma_{\mu\nu}$
is diagonal, but $\sigma_{xx} \ne \sigma_{yy}$. As a
result of the distortion of the Fermi contour, the current flow
is not parallel to the applied electric field except for the
special cases when the electric field is either parallel or
perpendicular to the in-plane magnetic field.

We finally comment on whether or not the transverse Hall field 
proposed by St\v{r}eda et al.\cite{Streda} is a possible
mechanism for the positive magnetoresistance. The nonequilibrium
distribution function in (\ref{disfn}) implies a spatial
redistribution of the electronic charge in a direction normal to
the interface given by\cite{tang2}
\begin{equation}
\delta n(z) = 2\sum_{n{\bf k}} g_n({\bf k}) |\phi_{nk_y}(z)|^2
\end{equation}
where $\phi_{nk_y}(z)$ is the subband wavefunction in the
presence of the in-plane magnetic field. 
This redistribution gives rise to a transverse Hall field and a Hall
potential which is estimated to be of the order of
microvolts under typical current carrying 
conditions.\cite{stern,tang2} These potentials are small on the
energy scale of the confining potential\cite{footnote1}
and will lead to small changes in the energy band dispersion.
Corrections to the Boltzmann equation (for
example, as a result of changes in the electron velocity) are
therefore of higher order in the applied electric field and, at
the level of the linearized Boltzmann theory, can be safely neglected.
A transverse Hall field is therefore
not relevant to in-plane transport.

\subsection{Evaluation of the Scattering Elements 
$P_{n n'} ({\bf k}, {\bf k}')$}

We now turn our attention to the calculation of 
$P_{n n'}({\bf k}, {\bf k}')$ in (\ref{scatrate}) for a
situation in which scattering is due to remote ionized impurities.  
Using the form of the subband states in the presence of a
parallel magnetic field, the potential matrix element has the form
\begin{equation}
\langle n{\bf k} \vert V({\bf r}) \vert n'{\bf k}' \rangle
={1\over{A}} \int dz \phi_{nk_y}(z) \phi_{n'k_y'} (z) 
V(z,{\bf q})
\label{mata}
\end{equation}
where ${\bf q}={\bf k} - {\bf k}'$ and
\begin{equation}
V(z,{\bf q})=
\int d{\bf s}
e^{-i{\bf q} \cdot {\bf s}} V({\bf s}, z)
\end{equation}
is the 2D Fourier transform of the scattering potential.
We note that the evaluation of (\ref{mata}) for the parallel field
configuration is considerably more involved because of the
$k_y$-dependence of the subband states.  As a result, the potential
matrix elements do not simply depend on the momentum transfer ${\bf q}$
which appears in the Fourier transform of the impurity potential.

We assume that the impurity potential arises from a distribution of
remote ionized impurities located at sites
$({\bf R}_i, Z_i)$.  These impurities (of charge $e$) give rise to the
bare electrostatic potential
\begin{eqnarray}
\phi^{ext}(z,{\bf q}) &=& \sum_i \left ( {{2 \pi e}\over{\kappa q}}
\right ) \; e^{-qz} \; e^{qZ_i} \; e^{-i {\bf q} \cdot {\bf R}_i}
\nonumber \\
&\equiv& \left ( {{2 \pi e}\over{\kappa q}} \right )
I({\bf q}) e^{-qz}
\label{start}
\end{eqnarray}
where a two dimensional Fourier transform has been taken in the 
plane of the 2DEG and we have assumed that the position of the
charged impurities is such that $Z_i < z$. 
$\kappa$ is the dielectric constant of the material in
which the 2DEG is imbedded.
The quantity $I({\bf q})$ determines the amplitude of
the exponentially decaying Fourier transform and
contains all information regarding the spatial
distribution of the ionized donor impurities.  

The external impurity potential is screened by the 2DEG and it 
is the final screened potential which is responsible for the 
electron scattering. We shall account for the 
screening at the level of the random phase approximation in
which the electrons respond self-consistently to the ionized impurity
potential.  Defining $\chi^0(z, z', {\bf q})$ to be the 2D Fourier
transform of the independent particle density response function, the
change in electron density due to the impurities is given by
\begin{equation}
\delta n(z,{\bf q}) = e \int dz' \chi^0 (z, z', {\bf q})  \phi^{tot}
(z',{\bf q})
\end{equation}
where $\phi^{tot} (z,{\bf q})$ is the total electrostatic potential
acting on the electrons.  This potential is given by
\begin{equation}
\phi^{tot} (z,{\bf q}) = \phi^{ext}(z,{\bf q})+\phi^{ind}(z,{\bf q})
\end{equation}
where the second term is the induced electrostatic potential
\begin{equation}
\phi^{ind} (z,{\bf q}) = - \left ( {{2 \pi e}\over{\kappa q}} \right )
\int dz' e^{-q \vert z-z' \vert} \delta n (z',{\bf q})
\label{indpot}
\end{equation}
arising from the induced electron charge density.

In the absence of a parallel magnetic field, the subband envelope
functions are ${\bf k}$-independent and the calculation of the density
response function is relatively straightforward.  It takes the form
\begin{equation}
\chi^0(z, z', {\bf q}) = \sum_{\beta} f_{\beta}(z) f_{\beta}(z')
\chi^0_{\beta}(q)
\label{screen}
\end{equation}
where the index $\beta$ represents a pair of subband indices
$(n,n')$ and $f_{\beta}(z)$ is the product $\phi_n(z) \phi_{n'}(z)$ 
of two subband envelope functions.  The factor $\chi_{\beta}^0(q)$
is just the intersubband response function for an ideal 2DEG which
depends on the subband structure through the subband energy levels and
the position of the Fermi level.  Analytic expressions for
$\chi_{\beta}^0 (q)$ are available (see, for example, Ref.\
\onlinecite{fletch2}).

Once a parallel magnetic field is present, the subband states acquire 
a non-trivial $k_y$ dependence and $\chi^0(z, z', {\bf q})$ 
cannot be obtained analytically.  To avoid an excessive and
largely unnecessary amount of
numerical work, we shall make a simple approximation 
motivated by
the following observation.  Although the individual subband states are
strongly modified by the parallel field, depending on the value of
$k_y$, the overall electron density distribution is relatively
insensitive to the field.  This is illustrated 
for Sample 2 in Figure~2, where we
compare the ground state density for $B_{\parallel}=0T$ to the density
at the relatively high field of $B_{\parallel}=10T$.  As can be seen,
the effect of the field is minor, even for this case in which
the second occupied subband is depopulated by the magnetic
field.  One would therefore expect the screening of the
impurity potential to take place in the presence of the parallel 
field in much the same way as in the
zero-field limit.  We therefore adopt the physically reasonable
approximation of screening the impurities by the zero-field response
function given by (\ref{screen}).

However, one complication must be addressed: higher occupied
subbands will depopulate with increasing parallel magnetic field.
Since different subbands screen differently, maintaining
the zero field populations in the calculation of 
$\chi^0(z, z', {\bf q})$ 
introduces an error which can be avoided in the following way.  
The subband populations $n_i(B)$ are first determined from a fully
self-consistent electronic structure calculation which is known
to reproduce the observed field dependence quite 
accurately.\cite{mine}
We then make use of this information in the calculation of
$\chi^0_{\beta}(q)$ by simply defining fictitious subband 
energies $E_i^0$ to ensure that the subband densities
are given correctly, that is,
\begin{equation}
2 \pi n_i(B_{\parallel}) = E_F - E_i^0\,.
\label{subden}
\end{equation}
This is sufficient to define the intersubband response function
for the occupied subbands. In the cases we have dealt with, a
maximum of two subbands are occupied and we have therefore
truncated $\chi^0_{\beta}(q)$ to a two-by-two matrix
corresponding to the lowest two subbands. Once the second
subband is depopulated, the second subband still contributes
to the screening by providing final states to which the first
subband electrons can be excited. In this situation the second
subband is positioned relative to the Fermi level according to
the self-consistent calculation, while the first subband
energy is still determined by (\ref{subden}).
In practice we have found that these refinements have only a
slight effect on the calculated scattering matrix elements since
the lowest subband holds most of the electrons and contributes
most of the screening.
Nevertheless, since relatively little additional effort is
required to determine these corrections, we have retained them
in all of the screening calculations.

With this prescription for $\chi^0_{\beta}(q)$ and 
$\chi^0(z, z', {\bf q})$,
the screened impurity potential is obtained from the set of
equations (\ref{start})-(\ref{indpot}).
Expressing $\phi^{tot}({\bf q},z)$ as
\begin{equation}
\phi^{tot}(z, {\bf q})= \left ( {{2 \pi e}\over{\kappa q}} \right ) 
\; I({\bf q}) \; J(z,q)\,,
\end{equation}
we have
\begin{equation}
J(z,q)=e^{-qz} - \left ( {{2 \pi e^2}\over{\kappa q}} \right )
\sum_{\beta} \chi^0_{\beta}( q)
\int dz' \int dz'' e^{-q \vert z-z' \vert}
f_{\beta}(z') f_{\beta}(z'') J(z'',q)\,.
\label{main}
\end{equation}
We note that this integral equation has a separable kernel and
it can therefore be reduced to a matrix problem.
Multiplying (\ref{main})
by $f_{\alpha}(z)$ and integrating over $z$ gives
\begin{equation}
J_{\alpha}(q)=\int dz f_{\alpha}(z) e^{-qz} - \sum_{\beta} 
F_{\alpha\beta}(q) \chi^0_{\beta}(q) J_{\beta}(q)
\label{mainmat}
\end{equation}
where
\begin{equation}
J_{\alpha}(q)=\int dz f_{\alpha}(z) J(z,q)
\end{equation}
and
\begin{equation}
F_{\alpha\beta}(q)= \left ( {{2 \pi e^2}\over{\kappa q}} \right )
\int dz \int dz' f_{\alpha}(z) f_{\beta}(z') e^{-q \vert z-z' \vert}
\end{equation}
is a Coulomb form factor.  As mentioned earlier, we retain
two subbands in our calculations, so that the screening effect
of the second subband is still present even when it is depopulated.

The solution of (\ref{mainmat}) is substituted into (\ref{main}) 
to obtain the
$z$-dependence of the screened impurity potential required in the
calculation of the scattering matrix elements.  In particular,
the angular transition rate takes the form
\begin{equation}
P_{n n'}({\bf k}, {\bf k}') = {{m^*}\over{\hbar^3}}
\Big\arrowvert \left ( {{2 \pi e^2}\over{\kappa q}} \right )
\int dz \phi_{nk_y} (z) \phi_{n'k_y'} (z) J(z,q)
\Big\arrowvert^2 {1\over A}\overline{\vert I({\bf q}) \vert^2}
\label{transition}
\end{equation}
where the bar over the final factor denotes a configuration
average. For uncorrelated impurity positions, we have
\begin{equation}
{1\over A}\overline{\vert I({\bf q}) \vert^2} = 
\int dz \; \rho_i(z) \; e^{2qz}
\label{form}
\end{equation}
where $\rho_i(z)$ is the average spatial density of ionized impurities.
For the special case of a delta-doped layer with $\rho_i(z) =
n_d \delta(z+s)$, (\ref{form}) becomes $n_d e^{-2qs}$, showing
that the effect of the impurities diminishes exponentially with
the setback distance $s$.

We can compare this result with the idealized situation of
short range impurity scattering considered by Tang and
Butcher\cite{tang1}. For a sheet of $\delta$-function scatterers
located in the plane $z = z_0$, we find $P_{n n'}({\bf k}, {\bf
k}') \propto |\phi_{nk_y} (z_0)|^2 |\phi_{n'k_y'} (z_0)|^2$,
which must be integrated over $z_0$ to correspond to a uniform
distribution of scatterers throughout the region of the 2DEG.
This scattering rate is independent of $k_x$ and $k_x'$ which,
as noted by Tang and Butcher\cite{tang1}, simplifies the
solution of the Boltzmann equation in the case of ${\bf E}
\parallel {\bf B}_\parallel$. For the more realistic situation
of remote charged impurities this simplification does not arise
and the ${\bf E} \parallel {\bf B}_\parallel$ case requires a
treatment similar to that of the ${\bf E} \perp {\bf B}_\parallel$ case.

\section{Results and Discussion}

All of the interesting magnetotransport effects in the presence
of an in--plane magnetic field ultimately arise from the
distortion of the Fermi contours illustrated in Figure~1 and the
associated behaviour of the subband wave functions. That this
could lead to a positive transverse magnetoresistance was
already appreciated by Leadley et al. who viewed the electrons
on the distorted Fermi contour as different kinds of carriers
having different mobilities. 
The usual treatment of parallel transport\cite{ziman}
would then give rise to a positive magnetoresistance. 
They attributed variations in the mobility to
a ${\bf k}$-dependent effective mass which is due to
the altered band structure. However, within a Boltzmann
transport theory it is the electron velocity which emerges as
the important dynamical variable, and the effective mass appears
only when the energy dispersion is strictly parabolic, which is
not the case in the present situation. In any event, a much more
significant factor is the variation of the electronic scattering
rates around the Fermi contour. In the following we shall try
to differentiate between the kinematic
effects which arise from the energy band structure and the
dynamical effects associated with impurity scattering.

Figure~1 shows an example of the Fermi contour anisotropy
for a sample in which two subbands are occupied at zero field.
As can be seen, the distortion of the Fermi contour of the first
subband increases with parallel field strength and 
eventually takes on the shape of an egg. As this distortion is
developing, the energy separation between the first and second
subbands increases and leads to the depopulation of the second
subband. (For Sample~2, complete depopulation occurs
at approximately $B_{\parallel}\sim 5.6T$.) Interestingly, the
minimum in the second subband also displaces relative to that in
the first, so that the pocket of second subband states approaches
the first subband Fermi contour with increasing field. 
This too will be seen
to have an important effect. However, we shall begin by
considering the simpler situation in which only a single subband 
is occupied, either because the electron density is low or because
the field is sufficiently high to have depopulated the higher
subband. Since our calculations are done for zero temperature,
only the states at the Fermi energy are relevant and we can
restrict the solution of (\ref{transformed}) to $\epsilon=E_F$.
It should be understood that
all quantities are calculated at this energy.

One measure of the Fermi contour anisotropy is the Jacobian
defined in (\ref{jacobian}). In Figure~3 we plot the inverse of
this quantity as a function of the angular position around the
Fermi contour, together with the ratio of the magnitude of the
velocity to the free-electron velocity $\hbar k /m^*$. We
recall that the ${\bf k}$-vector in this context is defined 
with respect to the position of the subband minimum and not
the more usual ${\bf k}$-space origin. One consequence of this
definition is that there is only a small difference between the
curves in Figure~3, indicating that the normal to the Fermi
contour does not deviate much from the direction of ${\bf k}$. 
In this respect, the Fermi contour is still rather circular in 
nature, despite its visual appearance. On the other hand,
the four-fold variation of $J_1^{-1}(\phi)$ around the contour 
is showing that the modified energy dispersion is having a 
dramatic effect on the electron velocity. There are two effects
coming into play. The first is the overall elongation along the
$k_y$-axis which is a result of a magnetic field enhancement of
the effective mass in this direction. As discussed 
previously\cite{mine},
this leads to an enhancement of the electronic density of
states above the ideal 2D value. It is this effect that accounts
for the decrease in $J_1^{-1}(\phi)$ near $\phi=\pi$. However this
same effect is swamped near $\phi=0$ by a much larger reduction of
the velocity coming from the
flattening of the energy bands. Only at $\phi = \pi/2$ is there
no effect of the Fermi contour anisotropy, as the dependence of
the energy on $k_x$ is unchanged by the parallel magnetic field.

The difference in behaviour of the energy dispersion near $\phi=0$
and $\phi=\pi$ is a reflection of the
$k_y$--dependence of the electron wavefunctions.  Electrons
with $k_y$ negative ($\phi \rightarrow \pi$) experience a
magnetic potential which drives them into the interface, whereas
states with positive $k_y$ ($\phi \rightarrow 0$) are pushed
away from the interface.  The latter is a stronger effect,
since the confining potential in a direction away from the
interface is much softer, particularly when the background
acceptor density is low. These differences are illustrated in
Figure~4 which shows the centroid of the subband
probability densities as a function of $k_y$. One can
see that there is only a slight change in the centroid position as
the wavefunctions are driven into the interface ($\pi/2 < \phi <
\pi$), but there is a much larger variation for those states being 
pushed in the opposite direction ($0 < \phi <\pi/2$).  

The change in mean position has a dramatic effect on the 
angular scattering rate (\ref{transition}). It is particularly
sensitive to the wavefunction position
since the bare scattering potential is an exponential 
function of the distance from the impurity layer.  Furthermore,
those states which
are further from the interface experience the full screening
effect of the 2DEG which is interposed between them and the
ionized donor impurities.
The angular scattering rate
depends on the two angular variables $\phi$
and $\phi'$ which specify the orientation of the initial and
final wavevectors, respectively. Part of this angular dependence 
arises from the impurity form factor $\overline{\vert
I({\bf q}) \vert^2}$ which depends
on the momentum transfer ${\bf q} = {\bf k}'
- {\bf k}$. This factor is common to the zero field limit
and we therefore choose to plot a normalized
transition rate which has this factor removed in order to
isolate the new effects associated with the dependence of the
scattering matrix elements on the subband states. In particular,
for the case of a single subband we consider the quantity
\begin{equation}
\bar P_{11}(\phi, \Delta\phi) =
{{A} \over 
{\overline{\vert I({\bf q}) \vert^2}}} 
\; P_{11}(\phi, \phi+\Delta\phi).
\end{equation}
where $\Delta\phi$ is the angle through which the electron
starting at the point $\phi$ is scattered. 
Because of the anisotropy of the Fermi
contour, $P_{11}(\phi, \phi+\Delta\phi)$ is not a symmetric
function of the scattering angle, except at the special points
$\phi = 0$ and $\phi = \pi$. More generally, we have the
symmetry $P_{11}(\phi, \phi') = P_{11}(2\pi - \phi,
2\pi - \phi')$. 

Figure~5 shows the normalized scattering
rate as a function of the starting angle $\phi$ for several
scattering angles. For a given scattering angle $\Delta \phi$,
the momentum transfer ${\bf q}$ is approximately constant and
the dependence on $\phi$ is an indication of the strong anisotropy
arising mainly from the $k_y$-dependence of the subband 
wavefunctions. The curve for $\Delta
\phi = 0 $ corresponds to the limit of small angle scattering
and shows a strong maximum at $\phi = \pi$. At this angle, the
subband states are closest to the impurities where the
screened potential $J(z,q)$ is relatively large. As $\phi
\rightarrow 0$ (or $2\pi$), the subband states move away from the
impurities and the scattering rate diminishes accordingly. A
similar behaviour is seen for the other scattering angles,
although now an asymmetry with respect to $\pm\Delta\phi$ is
evident. The decreasing magnitude of the scattering rate with
increasing scattering angle is partly due to the momentum transfer
dependence of the screened potential $J(z,q)$ which decreases
with increasing $q$.

It is also of interest to consider the anisotropy of
the quantum lifetime defined in (\ref{lifetime}). Given that the 
scattering rate drops off rapidly with increasing momentum transfer 
${\bf q}$, the inverse quantum lifetime is dominated 
by small angle scattering.  In Figure~6 the anisotropy of the
quantum lifetime is illustrated for Sample~1 at a field of $10T$.
The lifetime shows a four-fold variation between its maximum at
$\phi = 0$ and its minimum at $\phi = \pi$, the latter occurring
when the angular scattering rate has its maximum, as shown in  
Figure~5. To isolate the effects of the shape of the Fermi contour 
itself,
we can perform a model calculation in which the actual subband
wavefunctions are replaced by their zero-field limit. 
The lifetime anisotropy in this case is shown by the dashed line 
in Figure~6. Since 
the angular scattering rate in (\ref{transition}) is now only
a function of the momentum transfer, its anisotropy is
relatively weak and the lifetime anisotropy is dominated by the
Jacobian factor in (\ref{lifetime2}). As a result, the dashed
curve in Figure~6 mimics the behaviour of $J_1^{-1}(\phi)$ in
Figure~3. It is clear from a comparison of the two curves in
Figure~6 that the full lifetime anisotropy is coming mainly from
the magnetic field dependence of the subband wavefunctions.
The variation with parallel magnetic field strength is shown in
Figure~7. The lifetime at zero field is of course isotropic and
the anisotropy about this value is seen to increase approximately
in proportion to the parallel field.

To illustrate the effect of multiple subband occupancy,
we have performed 
similar lifetime calculations for Sample~2 at an
intermediate field ($5T$) where both subbands are still occupied.  
In this instance,
we can define multi--band quantum lifetimes by the equation
\begin{equation}
\tau_{n n'}^{-1}(\phi) = \tau_{n n'}^{-1}({\bf k})=\sum_{{\bf k}'}
w_{n n'} ({\bf k}, {\bf k}')\,.
\end{equation}
These lifetimes represent the unweighted probability of an electron
scattering from the state ${\bf k}(\phi)$ in subband $n$ to all other
states in subband $n'$.  The two--band quantum 
lifetimes presented in Figure~8 can be seen to
exhibit quite  different anisotropies.
For $\tau_{11}$, we basically have the same behaviour shown in
Figure~7 for the single-subband sample, but the magnitude of the
anisotropy is significantly larger because of the smaller
acceptor density in this sample. $\tau_{22}$ has a similar
behaviour, although the anisotropy is reduced because of the 
smaller range of $k_y$ values in the second subband.
For $\tau_{21}$ we see a slight increase
in the lifetime as $\phi \rightarrow \pi$.  This arises since
the second subband states near $\phi = \pi$ tend to 
experience a slightly larger average momentum transfer when
scattering into the first subband than do the states near $\phi =
0$.  However, the anisotropy of this intersubband lifetime is 
again small due to the small size of the pocket of second subband 
states. The variation of $\tau_{12}$ is much more dramatic, 
covering almost three orders of magnitude. This intersubband
lifetime shows a pronounced minimum near $\phi \sim \pi/8$ 
which is the region where the first Fermi contour comes closest
to the second subband. The reduced average momentum transfer
for these states results in a scattering `hot--spot' at which
the lifetime is relatively small.

In the zero field limit, the single subband transport lifetime
differs from the quantum lifetime by the appearance of an
additional $(1-cos\phi)$ weighting factor in the integrand of
(\ref{lifetime}). In the presence of a parallel magnetic field, 
there unfortunately is no similar expression which can be used
to define a transport lifetime. Instead, one must deal directly
with the non--equilibrium distribution function as determined by
the solution of the Boltzmann equation. This distribution 
function can always be expressed in the form
\begin{equation}
\xi_n(\phi) \equiv  {\bf E} \cdot \vec \Lambda_n(\phi)
\end{equation}
which introduces the vector mean-free-path $\vec
\Lambda_n(\phi)$. In general, the mean-free-path is not
parallel to the velocity vector at the point $\phi$ and for this
reason the conventional definition of a transport lifetime is not 
appropriate. Nevertheless, for a given direction of the electric
field, it is possible to parameterize the mean-free-path in terms
of a lifetime. For example, for ${\bf E}=E\hat {\bf x}$ (i.e.,
parallel to the in-plane magnetic field) we can write
\begin{equation}
\Lambda_{nx}(\phi) \equiv v_{nx}(\phi) \tau^{tr}_{nx}(\phi)\,.
\label{tau_nx}
\end{equation}
Since the zeroes of $\xi_n(\phi)$ coincide by symmetry with
those of $v_{nx}(\phi)$ when the electric field is in the
$x$-direction, $\tau^{tr}_{nx}(\phi)$ is a well-defined
quantity. This is not the case if the electric
field is oriented in the $y$-direction and a similarly defined
lifetime would exhibit singularities at
certain points on the Fermi contour which have no
physical significance. 
We shall therefore simply use $\tau^{tr}_{nx}(\phi)$ as a convenient
parameterization in order to visualize the anisotropy
of the solution to the Boltzmann equation.

Figure~9 shows the transport lifetime of Sample~1 defined according 
to (\ref{tau_nx}), normalized by the zero field transport
lifetime, at three magnetic fields. For lower fields we
see a monotonic angular variation similar to that of the quantum
lifetime, with those states at $\phi\rightarrow 0$ having a 
50 percent greater lifetime than those at $\phi\rightarrow \pi$.  
The states near $\phi = \pi/2$ which contribute the most to the
current have a lifetime very similar to the zero field lifetime.
As the field strength increases, the transport anisotropy changes
in character, with additional structure appearing near $\phi =
0$. More importantly, the average lifetime decreases with
increasing field and results in a positive parallel field
magnetoresistance.  This effect is illustrated in Figure~10
which shows the magnetoresistance for the current
parallel and perpendicular to the in--plane magnetic field.
The magnetoresistance is approximately parabolic and of a
sizable magnitude
for field strengths of the order of 10$T$. In addition,
we see that the magnetoresistance is larger for ${\bf J} \perp
{\bf B}_\parallel$ than for ${\bf J} \parallel {\bf B}_\parallel$,
in qualitative agreement with experiment\cite{Streda}, although
it should be noted that the experimental anisotropy seems to
vary from sample to sample in an unpredictable way.

We finally consider the transverse magnetoresistance which
arises with a perpendicular component of the magnetic field.
This component induces a cyclotron motion of the electrons
around the Fermi contour and leads to an averaging of the
anisotropic scattering rates. Figure~11 illustrates the behaviour
found for a series of parallel magnetic field strengths for a
situation in which two subbands are initially occupied. The
model parameters in this case are
$n_A=0.3\times 10^{11} cm^{-2}$ and $n_D=7.4\times 10^{11} cm^{-2}$
($n_T$ is still $7.1\times 10^{11} cm^{-2}$) instead of the
values $n_A=0.1\times 10^{11} cm^{-2}$ and $n_D=7.2\times
10^{11} cm^{-2}$ used for Sample~2.  This adjustment was
made since the two subband mobilities  for Sample~2
are almost identical and no positive
magnetoresistance is found for zero parallel field. 
By increasing the acceptor concentration the calculated 
magnetoresistance correponds more closely to that observed 
experimentally and a more meaningful comparison of the parallel
field dependence can be made. In addition, with the adjusted
parameters the second subband depopulates 
at approximately $4T$, which is in better agreement with the
G590c sample of Leadley et al.\cite{leadley}

The $B_{\parallel}=0T$ curve in Figure~11 is the usual positive
magnetoresistance associated with the different carriers in the
two subbands (although it should be stressed that the carriers
are {\it not} independent since intersubband scattering is
included\cite{zaremba}). 
As the parallel magnetic field is increased, both the depth 
of the magnetoresistance dip, and the field at which saturation
occurs, increase. This is in good qualitative agreement with the
observations of Leadley et al.\cite{leadley} on their G590c sample, 
although the saturation field we
find is approximately twice the value observed. Since saturation
occurs when $\omega_c\tau \simeq 1$, the discrepancy can be
explained in terms of the difference between 
our calculated mobility of $\mu=37.7 \; m^2/Vs$ and the experimental
value of $94.0 \; m^2/Vs$.
Once the depopulation field of about $4T$ is exceeded, we find
that the saturation field decreases suddenly to a value 
below the zero parallel field value due to the elimination of
intersubband scattering and then remains constant, in marked 
contrast to the behaviour found below the depopulation field.
The increase of the magnetoresistance with increasing parallel 
field in the single-subband limit is simply a consequence of
the increasing lifetime anisotropy shown in Figure~7. 
This behaviour is 
in excellent qualitative agreement with that observed.

In Figure 12 we give an example of the kind of anisotropy that
could be expected in the perpendicular field magnetoresistance.
Figure (a) is for the same sample discussed in Figure 10 and
shows a magnetoresistance which is only slightly larger for 
${\bf J} \parallel {\bf B}_\parallel$ than for ${\bf J} \perp
{\bf B}_\parallel$. Figure (b), corresponding to the $B_\parallel =
5T$ curve in Figure~11, shows a much larger anisotropy with the
magnetoresistance still being larger for
${\bf J} \parallel {\bf B}_\parallel$ than for ${\bf J}
\perp {\bf B}_\parallel$. This relative magnitude of the two
magnetoresistances persists over a range of sample parameters
and is opposite to the relative magnitude oberved in one
particular sample.\cite{Streda} We have no explanantion for
this difference. It should also be noted that the experimental
magnetoresistance tends not to saturate, but passes through
a maximum before falling at higher magnetic 
fields.\cite{leadley,Streda}

To further quantify our results we show in Figure~13 the
normalized magnetoresistance
\begin{equation}
{{\Delta \rho_{xx}}\over{\rho_0}} = 
{{\rho_{xx}(B_{\perp}=\infty)-\rho_{xx}(B_{\perp}=0T)}
\over{\rho_{xx}(B_{\perp}=0T)}}
\end{equation}
for a range of acceptor densities with the total electron
density fixed at $n_T=7.1\times 10^{11} cm^{-2}$.
A decrease in $n_A$ can be induced by illumination.
In agreement with observation, the magnetoresistance
increases with decreasing $n_A$; Figure~14(a) shows this dependence
in a different way.  This variation reflects the effect of 
illumination on the confining potential. As $n_A$ decreases, the
confining potential becomes softer and the subband states are more
strongly affected by a parallel magnetic field. 
In fact, we have found that $\Delta\rho_{xx}/\rho_0$ is far more
sensitive to variations in $n_A$ than in any of the other
material parameters. 
For example, Figure~14(b) illustrates the effects of varying the
total electron density $n_T$ while keeping the parallel magnetic field
constant.  For low values of $n_A$, an increase in the total electron
density translates into a reduction of the anisotropy in the system and
a lower $\Delta\rho_{xx}/\rho_0$.  This is mainly due to an increase in the
confinement of the self--consistent heterostructure potential
with increasing $n_T$. For
higher values of $n_A$, the acceptors are themselves providing a 
stronger confining potential, and the effect of varying
$n_T$ is greatly reduced.

A comparison of our results in Figure~13 with the experimental
data of Leadley et al. permits an estimate of the acceptor
densities in the experimental samples. 
We find that $n_A\simeq 0.35\times 10^{11} \; cm^{-2}$ 
is appropriate for the illuminated sample (G590c), while
$n_A\simeq 1.2 - 1.4\times 10^{11} \; cm^{-2}$ for the dark sample 
(G590a).  These values indicate that the change in acceptor
density as a result of illumination is 
$\Delta n_A \simeq 1\times 10^{11} \; cm^{-2}$, which is
consistent with the previously determined value\cite{fletch}.
Interestingly, our estimates suggest that a significant residual
acceptor density remains after illumination, in contrast to what
is sometimes assumed. If the acceptor density were eliminated
completely, our calculations would yield a much larger
magnetoresistance than observed.

\section{Conclusions}

We have performed extensive calculations of the transport properties
of
a 2DEG in the presence of parallel and perpendicular magnetic field
components. The in--plane component leads to a significant
perturbation of the electronic structure which manifests itself
as a distortion of the Fermi contours and a magnetic field
depopulation of higher lying subbands. These features are one
source of the anisotropy that emerges in both the quantum and
transport lifetimes. We have also analyzed the impact of the
parallel magnetic field on the scattering of electrons from
the long-range potential due to
remote ionized donors. The scattering at different
points on the Fermi contour is found to be highly anisotropic
as a result of the field dependence of the subband
wavefunctions. This detailed scattering information is then used
in a Boltzmann transport theory which treats the perpendicular
field component semiclassically. Our general solution of the
Boltzmann transport equation includes fully the effects of the
electronic structure, intersubband scattering and the anisotropy
of the nonequilibrium distribution function. Application of the
theory to a realistic heterojunction is found to yield results
which are in good qualitative agreement with experiment. In particular,
our calculations account for the observed positive transverse
magnetoresistance and its detailed dependence on the magnitude of the
in--plane magnetic field in both the one- and two-subband regimes.
It is clear from our calculations that a careful treatment of the
electronic structure, long-range impurity scattering and
transport behaviour are all needed to obtain a
complete understanding of the experimental results.

\section*{Acknowledegments}

This work was partially supported by grants from the Natural Sciences
and Engineering Research Council (NSERC) of Canada.

\begin{figure}
\caption{Constant Fermi energy contours for Sample~2 with a 
parallel magnetic field of (a) $5T$ (two-band occupancy) and (b) 
$10T$ (single-band occupancy).  $k_x$ and $k_y$ are in units of
inverse Bohr radii ($(a^*)^{-1}$).  The angular variable 
$\phi$ defines the position on the Fermi contour with respect to
the position of the first subband minimum, $k_1^{min}$;
a similar angular coordinate is defined for the
second subband, but is not shown.
}
\end{figure}

\begin{figure}
\caption{The spatial density profile of the 2DEG in Sample~2 as a
function of distance from the interface, for zero magnetic field
(solid line) and for a parallel magnetic field of $10T$
(dashed line).
}
\end{figure}

\begin{figure}
\caption{The inverse Jacobian
$m^* {\bf v}_n \cdot \hat{\bf k} / \hbar k$, along the Fermi contour 
of the first subband, for Sample~2 in the presence of a $10T$ 
parallel magnetic field (solid line).  Also included for comparison
is the function $m^* v_n / \hbar k$ (dashed line).
}
\end{figure}

\begin{figure}
\caption{The centroid of the first subband wave function relative 
to the interface, as a function of angular position on the Fermi
contour. The results are for Sample~2 in the presence of a 
$10T$ parallel magnetic field.
}
\end{figure}

\begin{figure}
\caption{The angular scattering rate 
$\bar P_{11}(\phi,\Delta\phi)$
of the first subband, for Sample~2 with a $10T$ parallel magnetic
field. The scattering rate is normalized by the small-angle
scattering rate at $\phi = \pi$.
The curves are labelled by the scattering angle
$\Delta\phi$, with the solid curves corresponding to positive 
scattering angles and the dashed curves to negative angles.
}
\end{figure}

\begin{figure}
\caption
{The first subband quantum lifetime, as a function of
the angular position on the Fermi contour, for Sample~1 in the 
presence of a $10T$ parallel magnetic field.  The solid line
is the full calculation using the $k_y$--dependent wavefunctions,
while the dashed line has been computed using only the constant 
zero field wavefunctions as described in the text.
}
\end{figure}

\begin{figure}
\caption{As in Figure 6, but for a range of parallel magnetic 
field strengths.
}
\end{figure}

\begin{figure}
\caption{The multiple-subband quantum lifetimes $\tau_{n n'}(\phi)$
for Sample~2 in the presence of a 5T parallel magnetic field, 
as a function of the angular coordinate.
}
\end{figure}

\begin{figure}
\caption{The effective transport lifetime of Sample~1 as a 
function of the angular coordinate, for an applied electric
field parallel to the in-plane magnetic field. The lifetime has been
normalized by the zero-field transport lifetime of the first subband.
The curves are labelled by the parallel magnetic field.
}
\end{figure}

\begin{figure}
\caption{Variation of the magnetoresistance as a function of the
in-plane magnetic field. The labels $xx$ and $yy$ correspond to
${\bf J} \parallel {\bf B}_\parallel$ and ${\bf J} \perp
{\bf B}_\parallel$, respectively.
}
\end{figure}

\begin{figure}
\caption{The transverse magnetoresistance for a series
of parallel magnetic field strengths. The parameters
characterizing this sample are described in the text. The second
subband depopulates at a parallel magnetic field of approximately 
$4T$. The curves have been offset arbitrarily for clarity, but
the scale of resistance variation is the same for each.
}
\end{figure}

\begin{figure}
\caption{The transverse magnetoresistance for ${\bf J} \parallel
{\bf B}_\parallel$ ($xx$) and ${\bf J} \perp {\bf B}_\parallel$
($yy$). (a) and (b) correspond to Samples 1 and 2,
respectively, both at a field $B_\parallel = 5T$.
}
\end{figure}

\begin{figure}
\caption{The normalized transverse magnetoresistance
as a function of parallel magnetic field strength. The different
curves are labelled by the acceptor density $n_A$, 
in units of $10^{11} \; cm^{-2}$. All samples have the same
electron density of $7.1 \times 10^{11} \; cm^{-2}$. The 
low-field termination of the curves occurs
at the point where the second subband becomes populated.
}
\end{figure}

\begin{figure}
\caption{(a) The normalized magnetoresistance as a function of $n_A$
at three different parallel magnetic fields,  
for the same set of samples as in Figure~11.
(b) As in (a), but for parallel magnetic field of $10T$ and
varying electron gas density, $n_T$, in units of $10^{11} \; cm^{-2}$.
}
\end{figure}


\begin{references}
\bibitem{footnote1} The redistribution of the electronic charge
must in principle be determined self-consistently, but this does
not alter the argument that the changes are proportional to the
applied electric field.
\end{references}
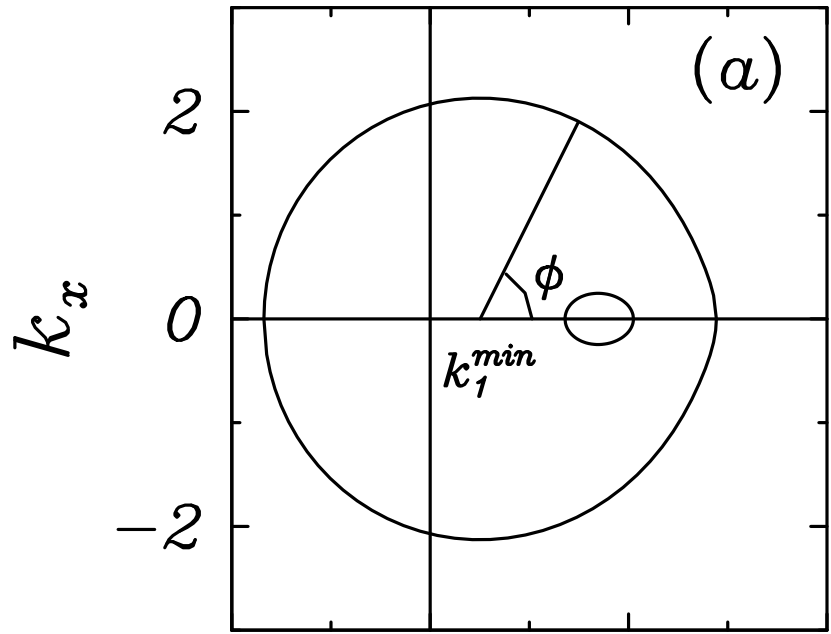
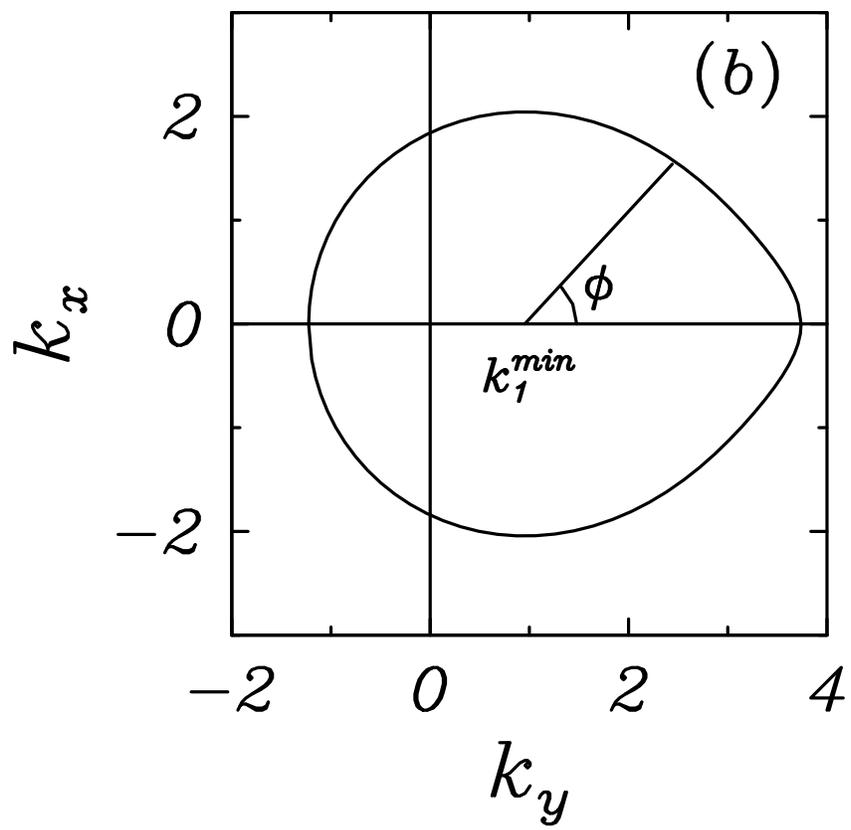
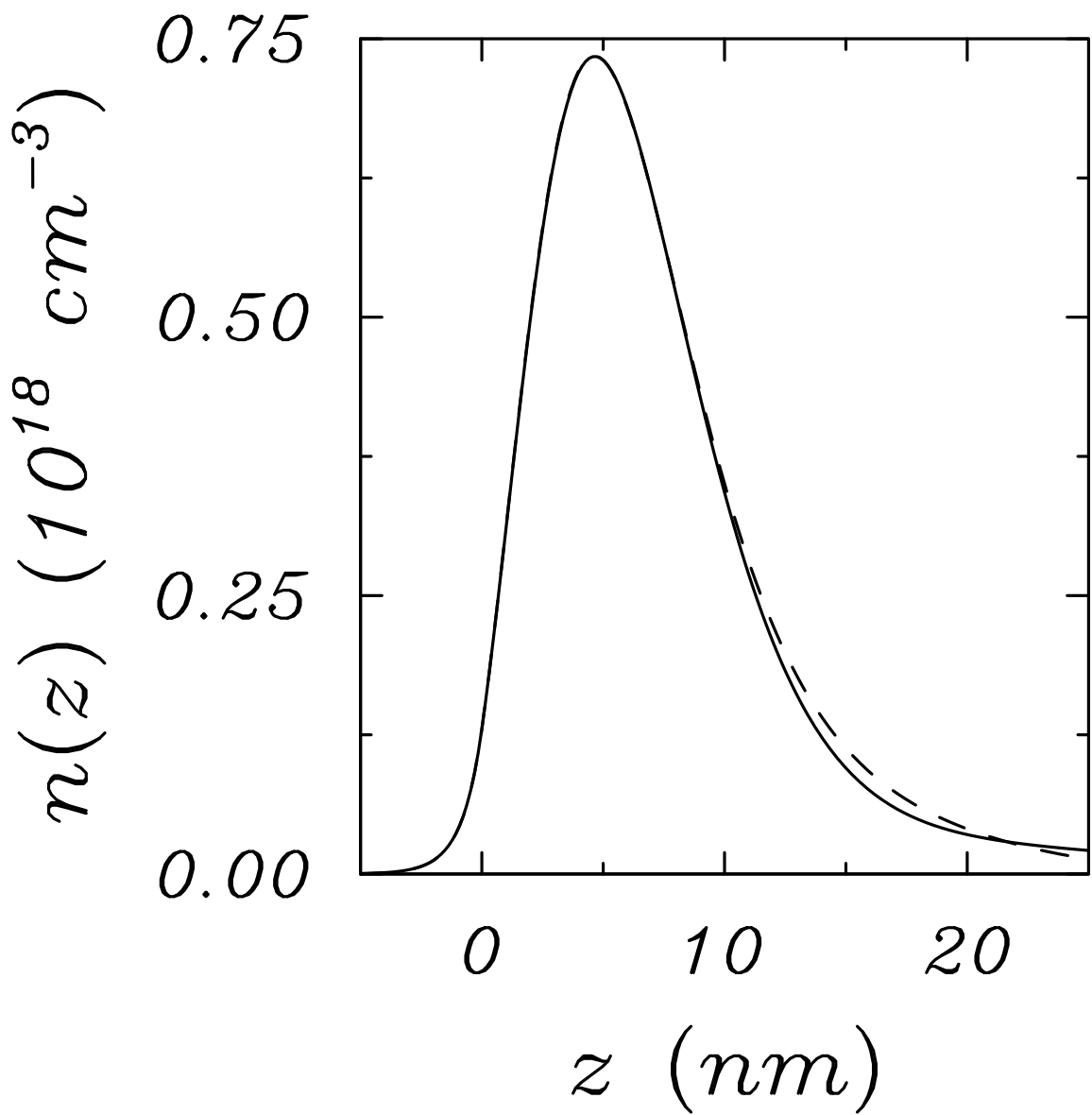
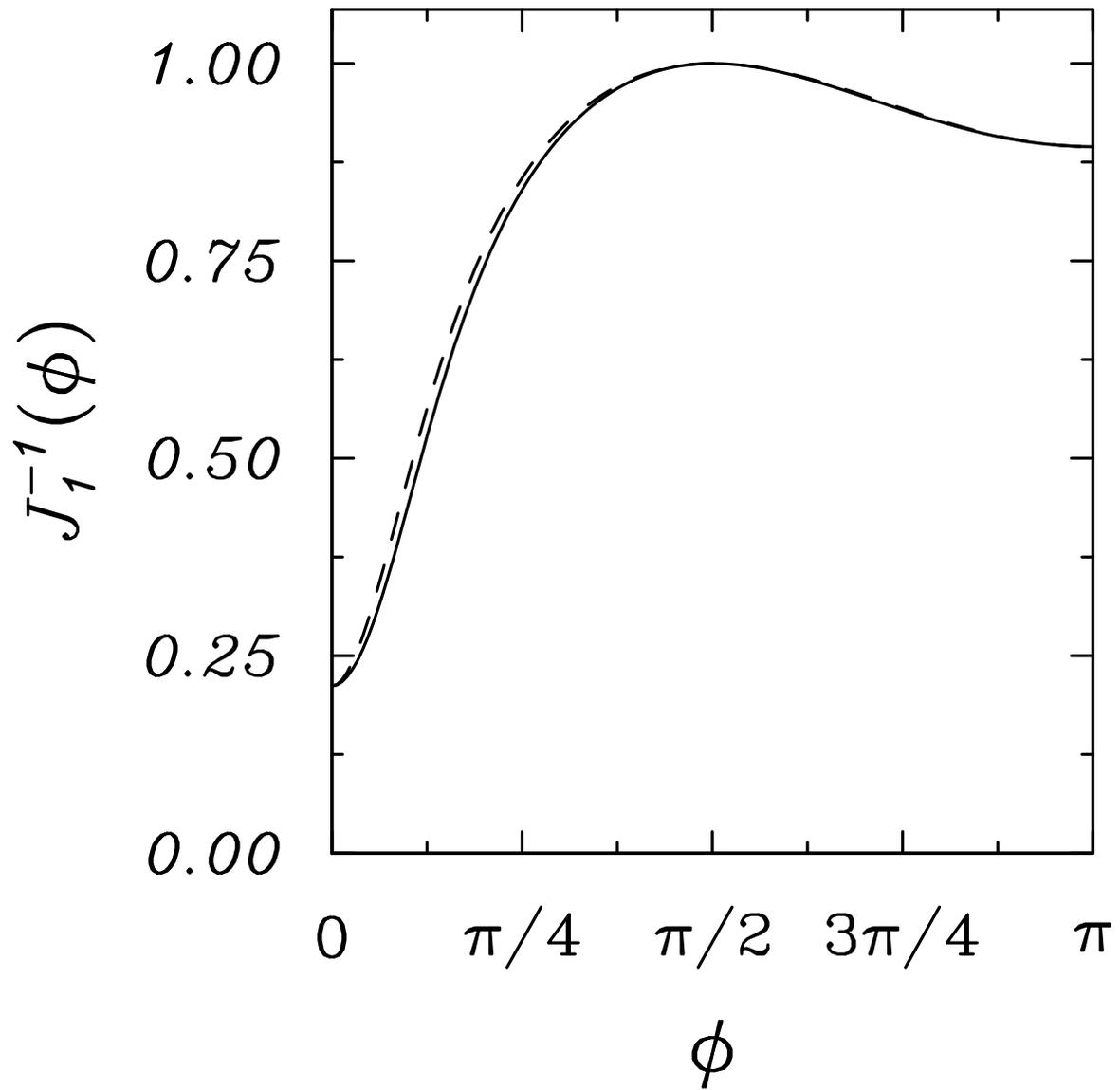
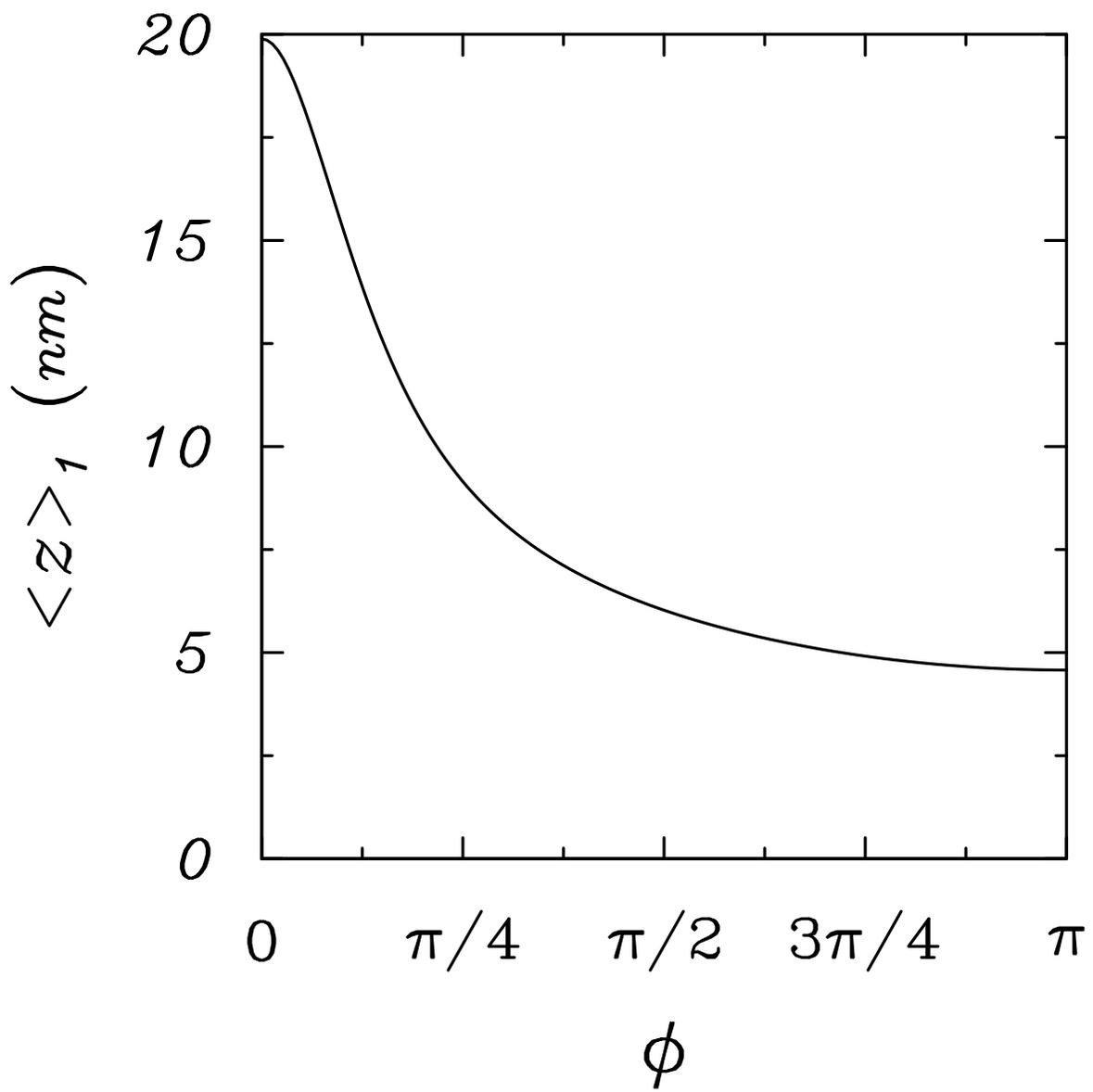
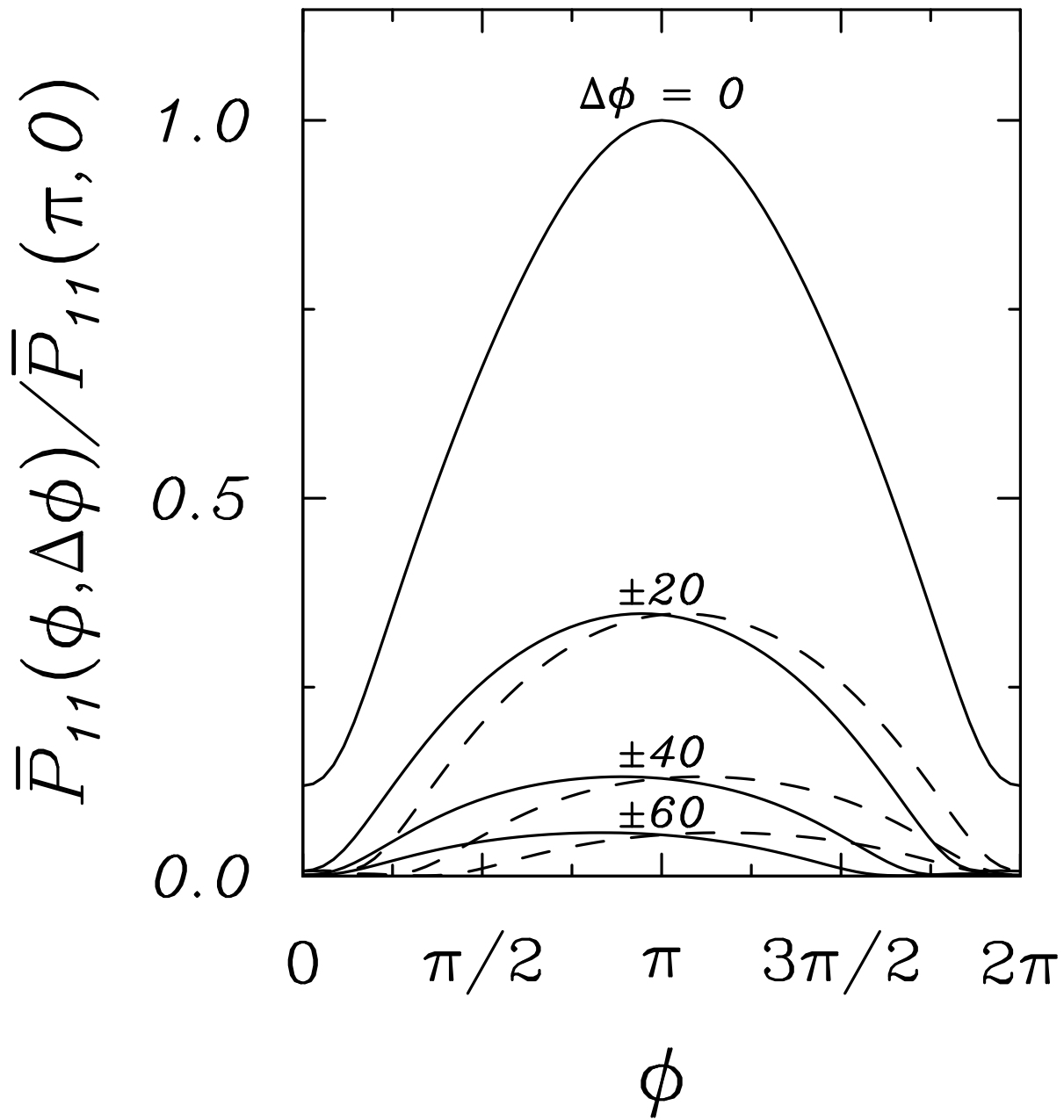
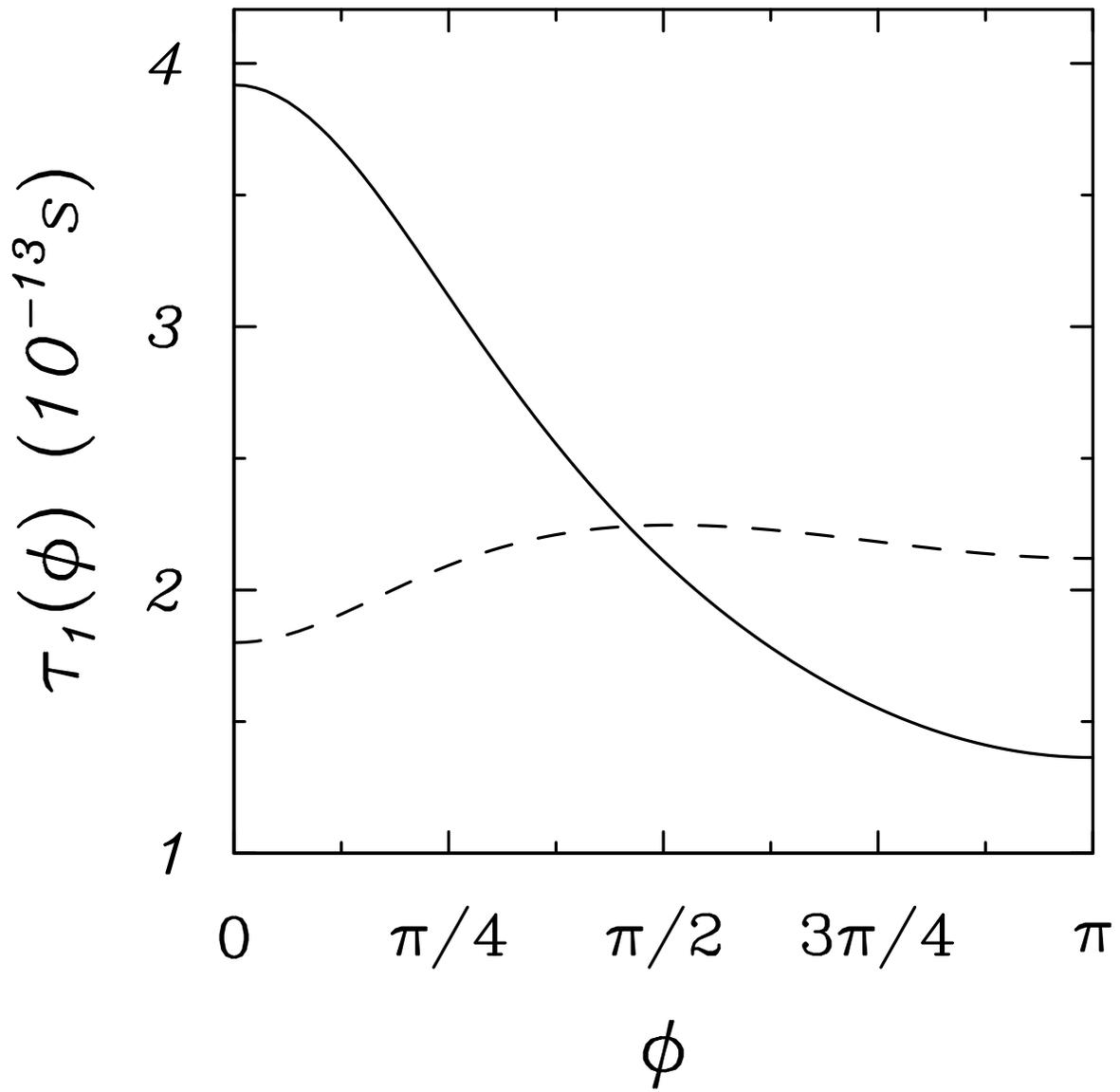
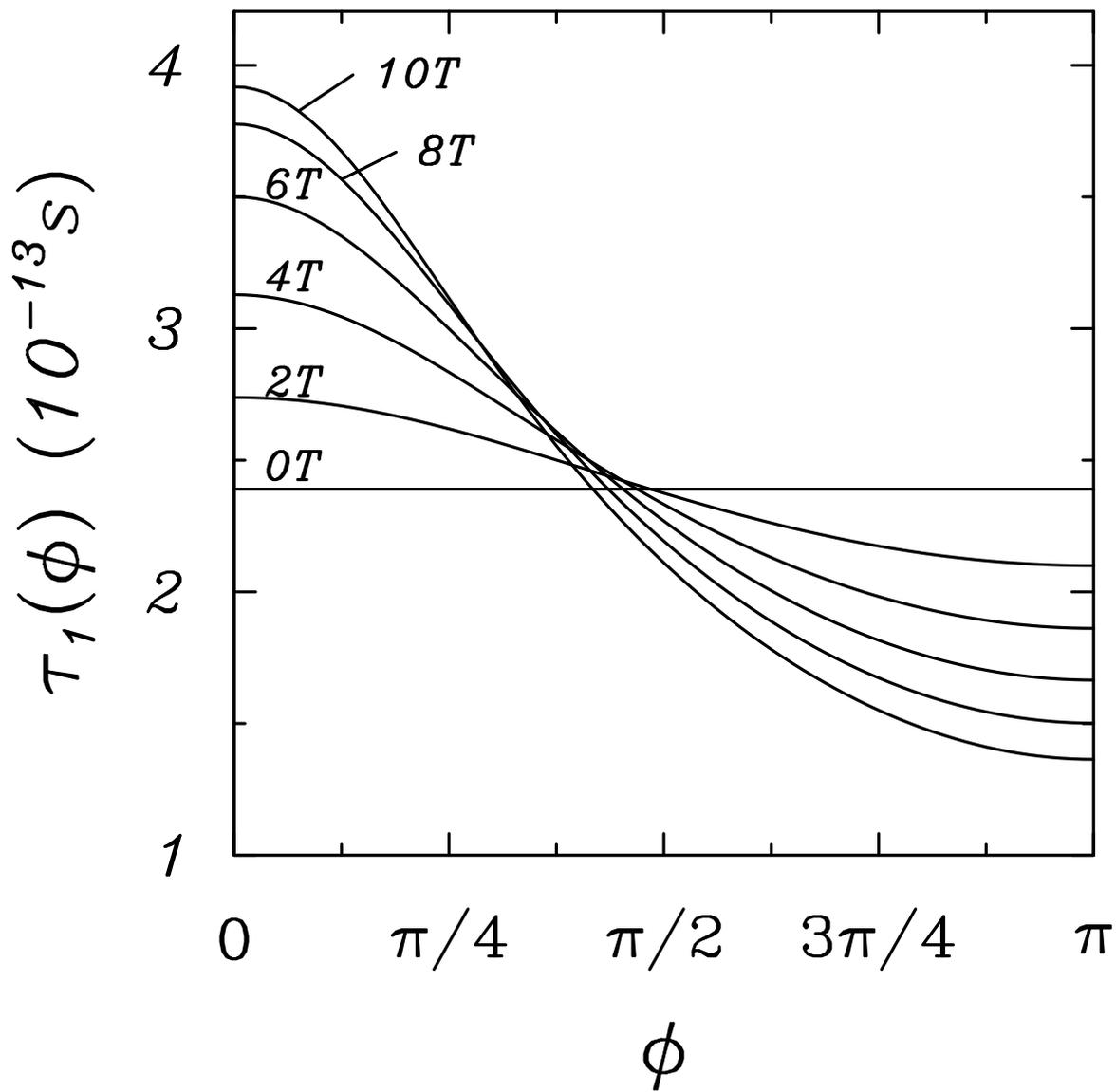
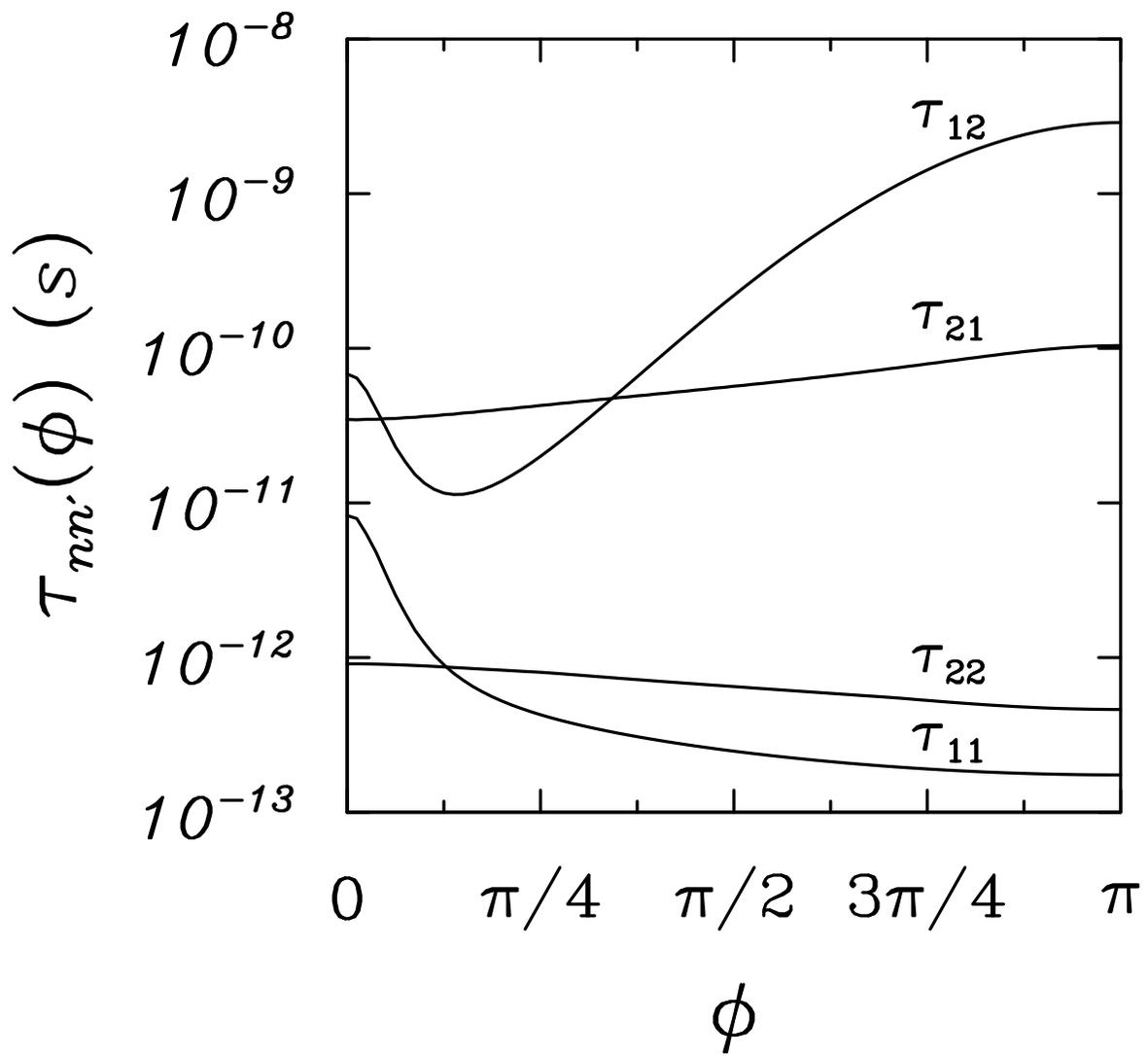
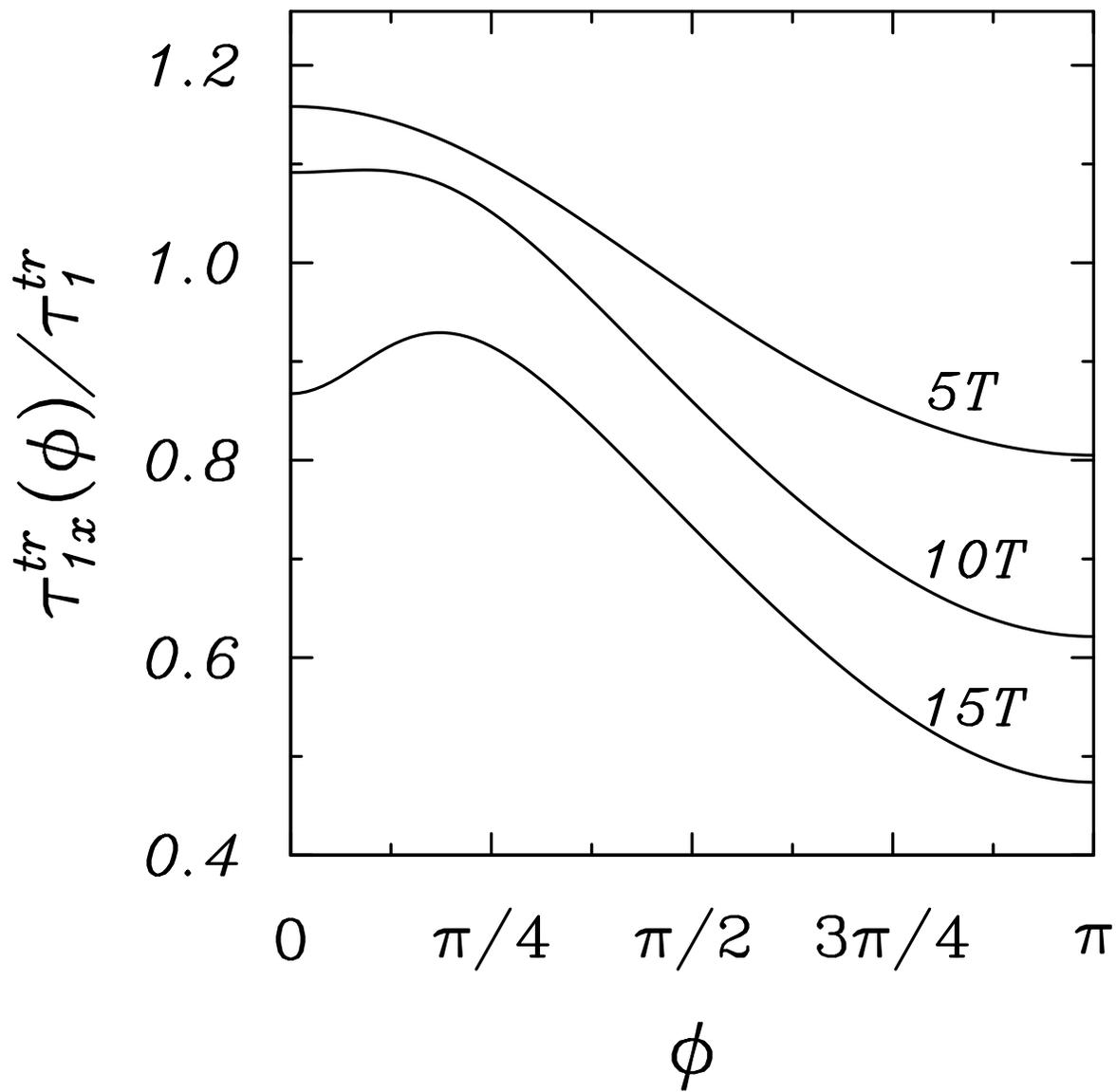
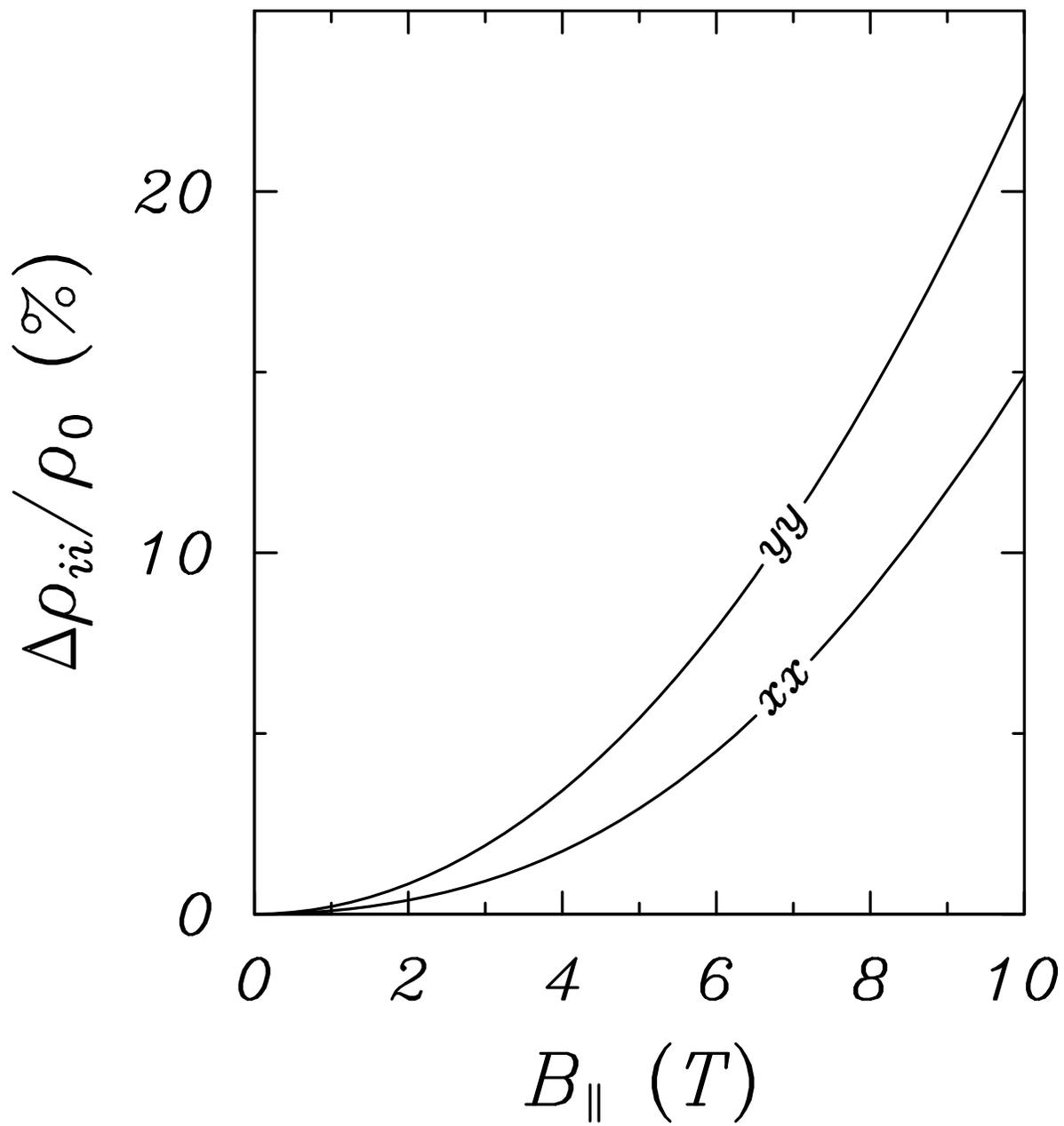
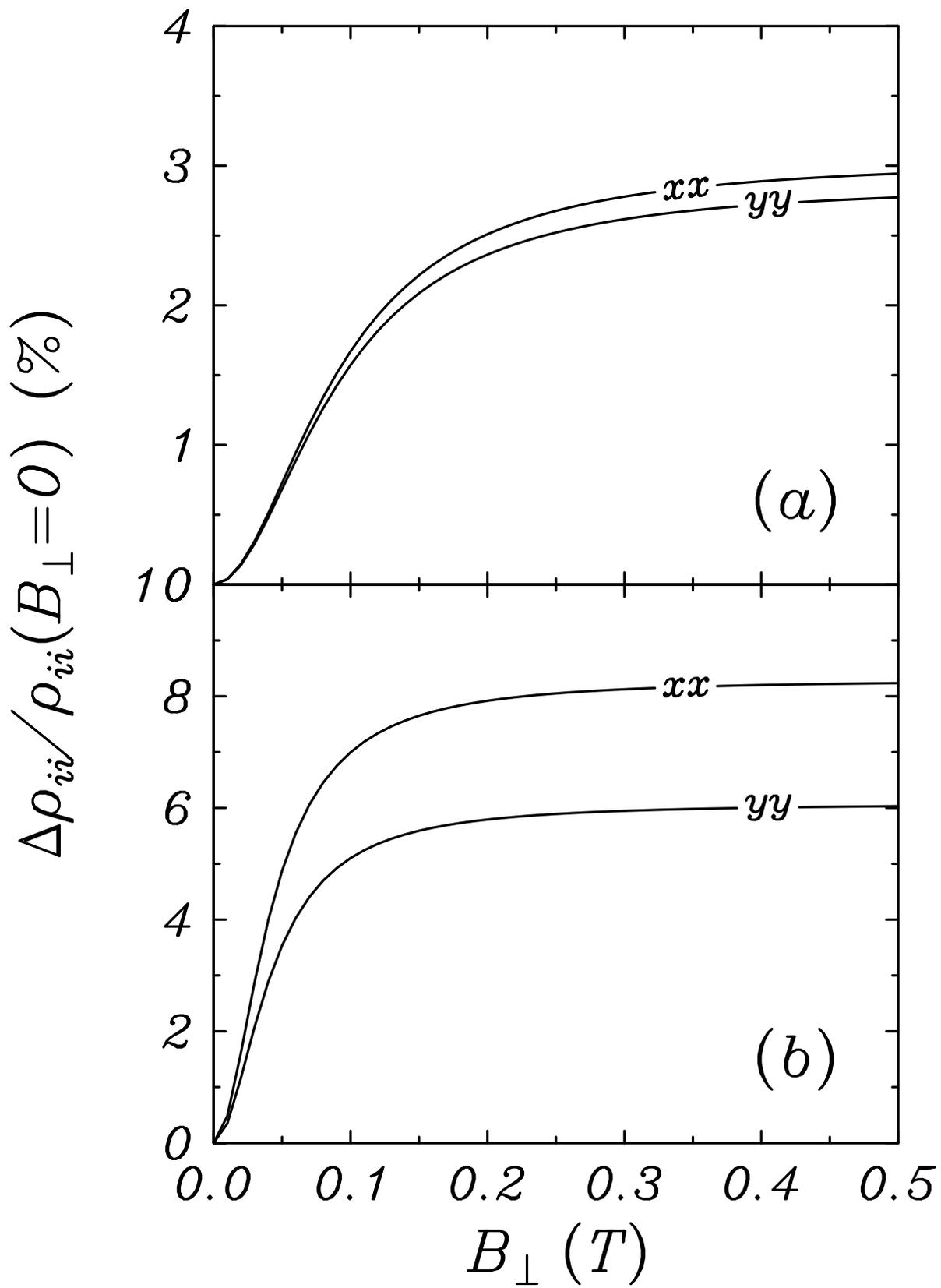
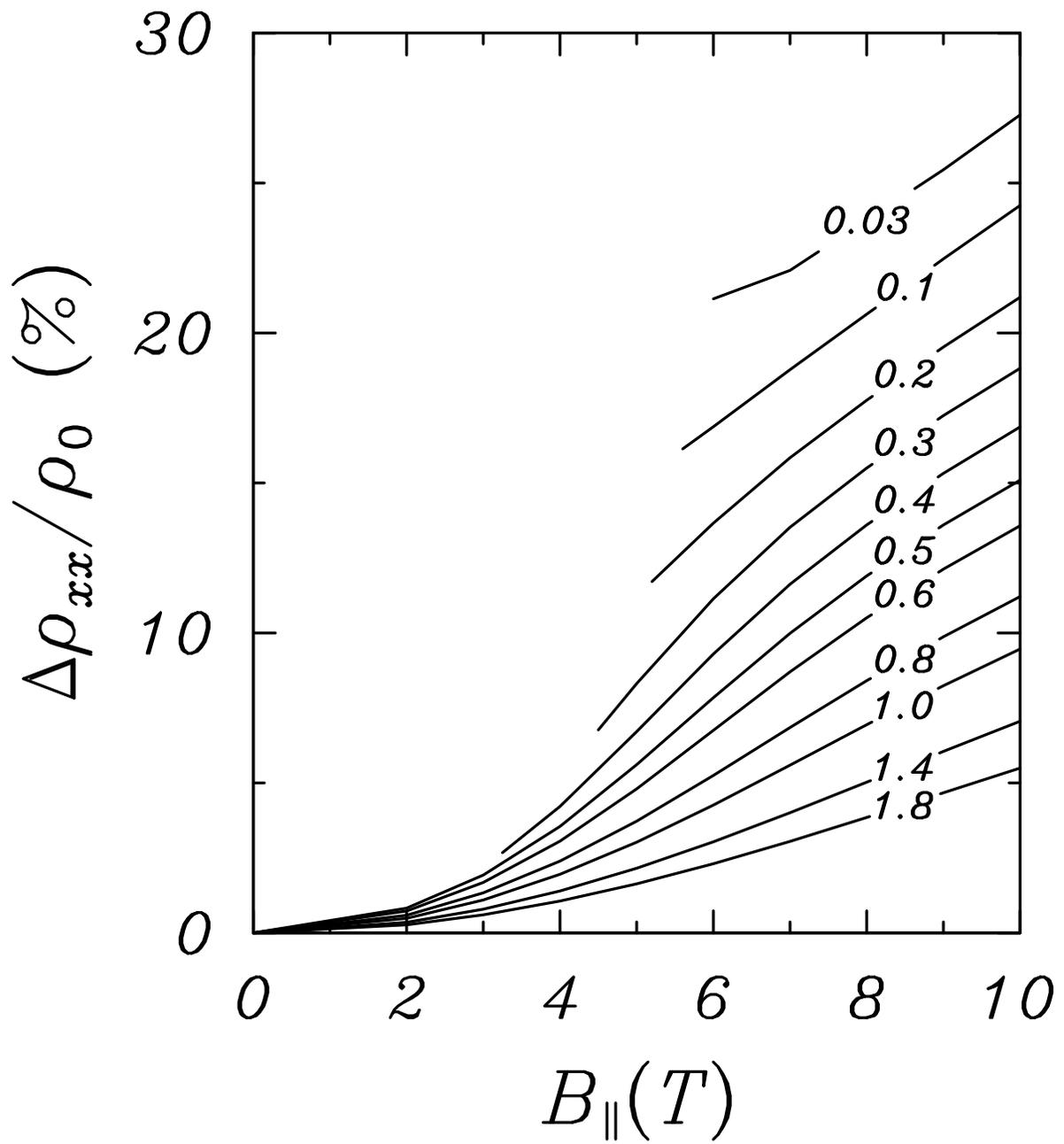
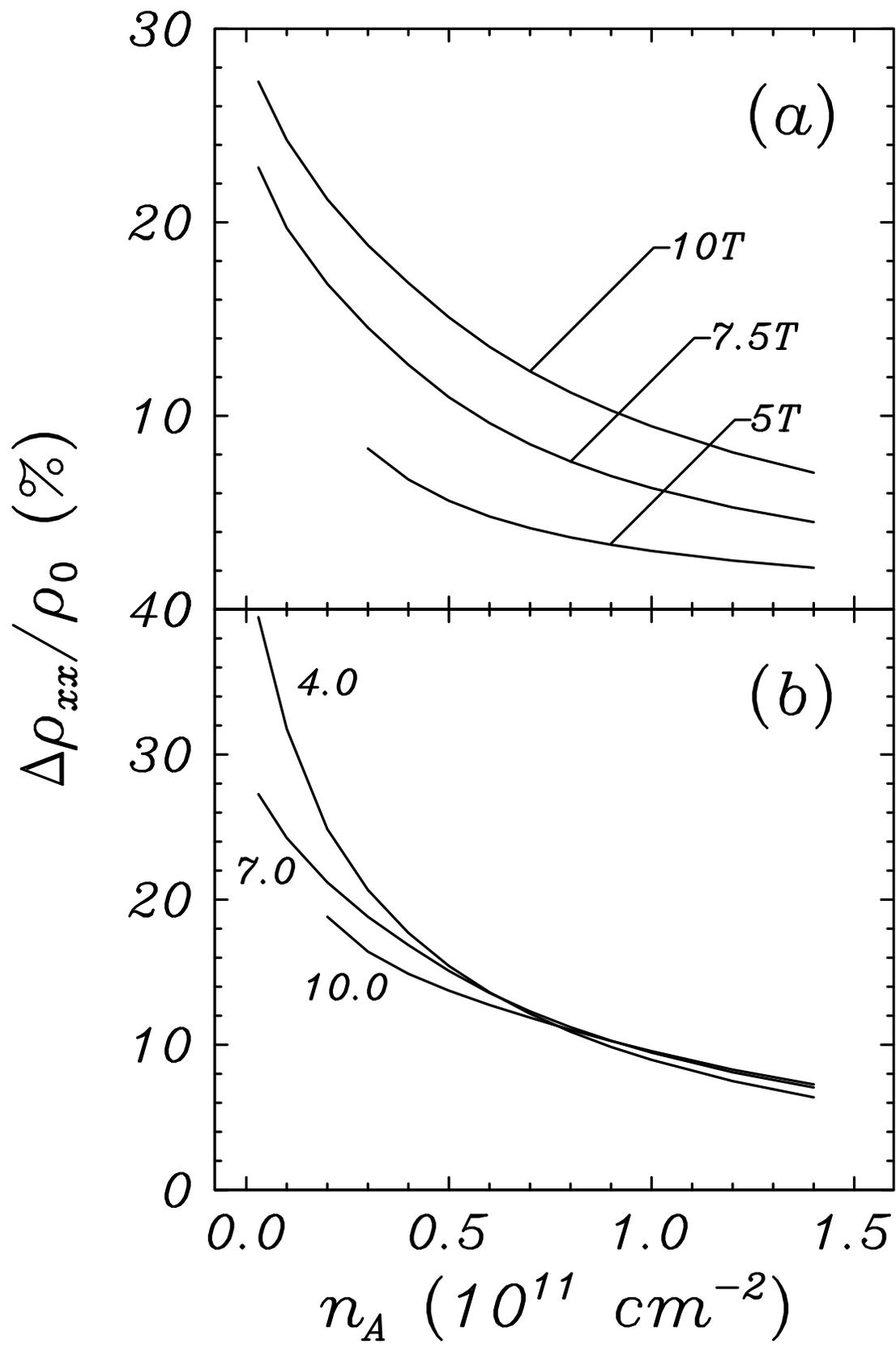
\end{document}